\documentclass[a4paper,fleqn,usenatbib]{mnras}

\usepackage{newtxtext,newtxmath}
\usepackage[T1]{fontenc}
\usepackage{ae,aecompl}
\usepackage{graphicx}
\usepackage{amsmath}
\usepackage{amssymb}

\usepackage{morefloats}

\newcommand\loiii{$L_{[\ion{O}{iii}]}$}
\newcommand\oiii{\ion{O}{iii}}
\newcommand\nii{\ion{N}{ii}}

\newcommand\re{$R_e$}

\newcommand\starlight{\textsc{starlight}}
\extrafloats{600}
\title[The first 62 AGN in MaNGA - II: resolved stellar populations]{The first 62 AGN observed with SDSS-IV MaNGA - II: resolved stellar populations}

\author[N. D. Mallmann et al.]{N\'icolas Dullius Mallmann$^{1,2}$,\thanks{E-mail: nicolas.mallmann@ufrgs.br}
	Rog\'erio Riffel$^{1,2}$,
	Thaisa Storchi-Bergmann$^{1,2}$,
	\newauthor 
	Sandro Barboza Rembold$^{2,3}$,
    Rogemar A. Riffel$^{2,3}$,
    Jaderson Schimoia$^{1,2,3}$,
	\newauthor 
	Luiz Nicolaci da Costa$^{2}$,
    Vladimir \'Avila-Reese$^{4}$,
    Sebastian F. Sanchez$^{4}$, 
	\newauthor 
    Alice D. Machado$^{2,3}$,
    Rafael Cirolini$^{2,3}$,
    Gabriele S. Ilha$^{2,3}$,
    Jana\'{i}na C. do Nascimento$^{1,2}$
\\
$^{1}$Departamento de Astronomia, Universidade Federal do Rio Grande do Sul - Av. Bento Gon\c calves 9500, Porto Alegre, RS, Brazil.
\\
$^{2}$Laborat\'orio Interinstitucional de e-Astronomia, Rua General Jos\'e Cristino, 77 Vasco da Gama, Rio de Janeiro, Brazil, 20921-400
\\
$^{3}$Departamento de F\'isica, Centro de Ci\^encias Naturais e Exatas, Universidade Federal de Santa Maria, 97105-900, Santa Maria, RS, Brazil
\\
$^{4}$Instituto de Astronom\'ia, Universidad Nacional Aut\'onoma de M\'exico, A. P. 70-264, C.P. 04510, M\'exico, D.F., Mexico
}

\date{Accepted XXX. Received YYY; in original form ZZZ}

\pubyear{2017}

\begin{document}
\label{firstpage}
\pagerange{\pageref{firstpage}--\pageref{lastpage}}
\maketitle

\begin{abstract}
We present spatially resolved stellar population age maps, average radial profiles and gradients for the first 62 Active Galactic Nuclei (AGN) observed with SDSS-IV MaNGA to study the effects of the active nuclei on the star formation history of the host galaxies. These results, derived using the {\sc starlight} code, are compared with a control sample of non-active galaxies matching the properties of the AGN hosts. We find that the fraction of young stellar populations (SP) in high-luminosity AGN is higher in the inner ($R$\ $\rm \leq 0.5\,$ $R_e$) regions when compared with the control sample; low-luminosity AGN, on the other hand, present very similar fractions of young stars to the control sample hosts for the entire studied range (1 $R_e$). The fraction of intermediate age SP of the AGN hosts increases outwards, with a clear enhancement when compared with the control sample. The inner region of the galaxies (AGN and control galaxies) presents a dominant old SP, whose fraction decreases outwards. We also compare our results (differences between AGN and control galaxies) for the early and late-type hosts and find no significant differences. In summary, our results suggest that the most luminous AGN seems to have been triggered by a recent supply of gas that has also triggered recent star formation (t\,$\leq$\,40\,Myrs) in the central region.
\end{abstract}



\section{Introduction} 

An important galaxy evolution stage is characterized by the Active Galactic Nuclei (AGN), a phenomenon that occurs when the galaxy's supermassive black hole (SMBH) is accreting matter from its surroundings, i.e., the accretion disk. Subsequent feedback processes start to happen, comprising radiation emitted by the hot gas in the accretion disc or by its corona, jets of relativistic particles, and winds emanating from outer regions of the disk.

Current models and simulations of gas inflows on tens to hundreds of parsec (pc) scales around galaxy nuclei lead to episodes of circumnuclear star formation \citep{Kormendy2013,Heckman2014,Zubovas2017}. \cite{Zubovas2017} suggests that there is a critical AGN luminosity in which the feedback of the nuclear activity increases the fragmentation of the gas clouds. Above this luminosity threshold, the feedback is powerful enough to remove the gas efficiently and stop fragmentation; for AGN luminosities under this threshold, however, the feedback is not efficient to compress the gas to high densities and enhance fragmentation. However, there is no consensus on whether AGN fueling occurs at the same time as the star formation \citep{Kawakatu2008}, or follows it during a post-starburst phase \citep{CidFernandes2005,Davies2007,Davies2009} or if it is not associated with any recent star formation \citep{Sarzi2007,Hicks2013}. 

A breakthrough in understanding the relation between the AGN and the surrounding stellar population can be reached by a simple, but thorough, investigation of whether young or intermediate age stars are present within few hundred pc of the AGN. If the youngest stellar types are present, AGN fueling is coeval with star formation; if instead intermediate age stars dominate the stellar population, fueling would be driven by a post-starburst and, thus the AGN phase would follow the starburst phase; finding only old stars would imply that gas inflow to the AGN is not necessarily linked to star formation.

Over the last few years, major observational effort to understand this co-evolution between AGN and the circumnuclear stellar population is being made using spatially resolved stellar population studies in large samples of galaxies \citep{Goddard2017,Zheng2017}. These studies, however, are not focused on comparing AGN hosts with non-active galaxies. One recent effort focusing on such kind of comparison was made by \citet{Sanchez2017} who found that AGN hosts are mostly morphologically early-type or early-spirals and that for a given morphology, AGN hosts are more massive, more compact, more centrally peaked, and rather pressure than rotationally-supported systems when compared to the non-active galaxies. However, these studies did not use a selected control sample of galaxies to match the fundamental properties of the AGN sample, nor considered the dependence on the AGN luminosities.

This is the second paper of a series in which we aim at studying the resolved stellar population as well as the gas emission properties of the AGN host galaxies observed with MaNGA and compare them with those of a control sample of non-active galaxies. In Paper I \citep{Rembold2017}, we have presented the AGN sample so far observed with MaNGA (available through the MPL-5) and have defined a control sample matching the AGN host galaxies in terms of galaxy masses, morphology, distance and inclination. In Paper I we have also characterized the stellar population properties of the AGN hosts as compared with those of the control sample for the single aperture SDSS-III spectrum that covers the inner 3" diameter nuclear region, using spectral synthesis via the {\sc starlight} program \citep{CidFernandes2005}.

Aimed at investigating the relation between the nuclear activity and the hosts' star formation history (SFH), in the present paper (Paper II), we use the MaNGA datacubes of the AGN and control sample defined in Paper I to obtain the resolved SFH and stellar population properties for these objects. These properties were compared between the AGN hosts and inactive galaxies in different luminosity ranges. This paper is organized as follows: brief description of the MaNGA subsample chosen for this work (Section 2); the method of stellar population synthesis as well as the base set of simple stellar populations (Section 3); the results of the synthesis for the AGN and control sample (Section 4); a discussion comparing the stellar populations of AGN and control galaxies (Section 5); and a conclusion in Section 6.

\section{Data}

The study of spatially resolved properties in galaxies were always undermined by the small sample size of past integral field spectroscopy surveys, not to mention the less numerous AGN. To address this problem, the Mapping Nearby Galaxies at Apache Point Observatory (MaNGA) survey \citep{Bundy2015} was developed to observe a large sample of nearby galaxies with integral field spectroscopy.

MaNGA is part of the fourth generation Sloan Digital Sky Survey (SDSS IV) along with APOGEE-2 \citep{Majewski2015} and eBOSS \citep{Dawson2016}. The survey aims to provide optical spectroscopy ($3600$\,{\AA}-$10400$\,{\AA}) of $\sim 10,000$ nearby galaxies (with $\langle z\rangle\,\approx\,0.03$). The observations are carried with fiber bundles of different sizes (19-127 fibers) covering a field of $12''$ to $32''$ in diameter. The selected sample is divided into ``primary'' and ``secondary'' targets, the former are observed up to $1.5$ effective radius (\re) whilst the latter is observed up to $2.5\,R_e$. For more details, see \citet{Drory2015,Law2015,Yan2015,Yan2016}.

The data used in the present work is a sub-sample of MaNGA data \citep[][MPL-5's Data Reduction Pipeline, DRP]{Law2016} selected in Paper I. In short, the AGN were selected from the MaNGA sample by crossmatching them with SDSS-III data products, using then the BPT diagram [\oiii]/H$\beta$ vs. [\nii]/H$\alpha$ \citep{Baldwin1981} to select the AGN. In addition, we have used the WHAN diagram (\citealt{CidFernandes2010}; \citealt{CidFernandes2011}) to eliminate from the AGN sample the ``LIERs", or ``fake AGN". The resulting AGN sample contains 62 objects. To study the relationship between AGN and the stellar populations of the host, we have chosen two control galaxies to match each of the selected AGN hosts. The matching was done according to the morphology (using concentration and asymmetry indices), axial ratios, redshifts, galaxy inclination, and total stellar masses. For more details regarding the AGN and control sample selection, see \cite{Rembold2017}.

\section{Stellar Population Synthesis}

We have used stellar population synthesis technique in order to derive the SFH of the galaxies of the AGN and control samples. We first briefly describe of the fitting code used. We then present a summary of the data preparation and processing pipeline we have developed to manage the fitting process.

\subsection{Fitting code} \label{fit}
To disentangle the contribution of each stellar population to the integrated spectra of each spaxel in the datacubes we employed the {\sc starlight} code \citep{CidFernandes2005}. In summary, this code combines the spectra of a base set of $N_{\star}$ template spectra $b_{j,\lambda}$ -- usually, simple stellar population (SSP) covering a range of ages and metallicities -- in order to reproduce the observed spectra $O_\lambda$. To generate the modeled spectra $M_\lambda$, the SSPs are normalized at an arbitrary $\lambda_0$ wavelength, reddened by the term $r_\lambda = 10^{-0.4 (A_\lambda - A_{\lambda_0})}$, weighted by the population vector $x_j$ (which represents the fractional contribution of the $j$th SSP to the light at the normalization wavelength $\lambda_0$), and convolved with a Gaussian distribution $G(v_\star, \sigma_\star)$ to account for the effects of velocity shifts in the central velocity $v_{\star}$ and velocity dispersion $\sigma_{\star}$. The model spectrum can be expressed as:

\begin{equation}
	M_\lambda = M_{\lambda_0} \left[ \sum_{n=1}^{N_\star} x_j\,b_{j,\lambda}\,r_\lambda \right] \otimes G(v_\star, \sigma_\star)
\end{equation}

\noindent where $M_{\lambda_0}$ is the synthetic flux at the wavelength $\lambda_0$. To find the best parameters for the fit, the code searches for the minimum of $\chi^2 = \sum_{\lambda_i}^{\lambda_f} [(O_\lambda - M_\lambda) \omega_\lambda]^2$, where $\omega_\lambda$ is the inverse of the error, using a simulated annealing plus Metropolis scheme. Further details on the code can be found in \citet{CidFernandes2005}.


\begin{figure*}
	\includegraphics[width=1.9\columnwidth]{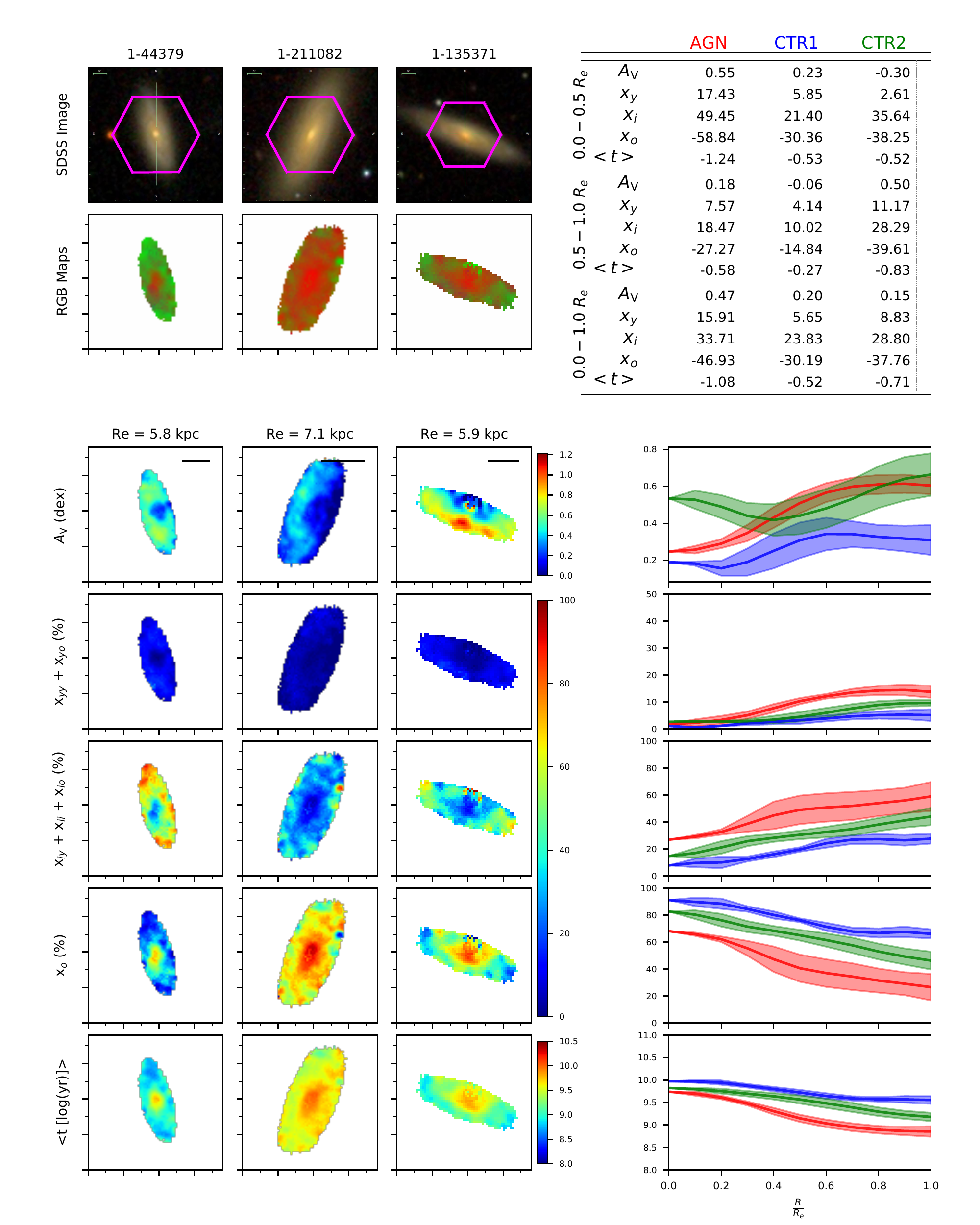}
    \caption{
    	Comparison between a late-type AGN and its control galaxies. {\bf Left side panels -} {\it Top set of panels:} SDSS image (the MaNGA field is indicated in magenta). Second row: composed RGB image using the binned population vectors [blue: young ($\rm X_Y$:  t$\leq$40\,Myr); green: intermediate age ($\rm X_I$: $ 40 < \leq < 2.5\,Gyr$); red: old ($\rm X_O$: $ t > 2.5\,Gyr $)]. {\it Bottom set of panels:} From top to bottom: visual extinction ($A_{\rm V}$), $\rm X_Y$, $\rm X_I$, $\rm X_O$ and mean age ($<t>$) maps. For display purposes we used tick marks separated by  5$"$. The solid horizontal line in the $A_{\rm V}$ maps represent 1\,$R_e$.
	{\bf Right side panels -} {\it Top:} summary table with the mean gradient values for each property in 3 different $R_e$ ranges.
    {\it Bottom:} average radial profiles, up to 1\,$R_e$, for AGN (red color) and control (blue and green colors). Shaded area represents $1\,\sigma$ standard deviation. 
	 For profiles smaller than 1\,$R_e$ the gradients were calculated using extrapolated values.
     }
    \label{fig:late-comp}
\end{figure*}

\begin{figure*}
	\includegraphics[width=1.9\columnwidth]{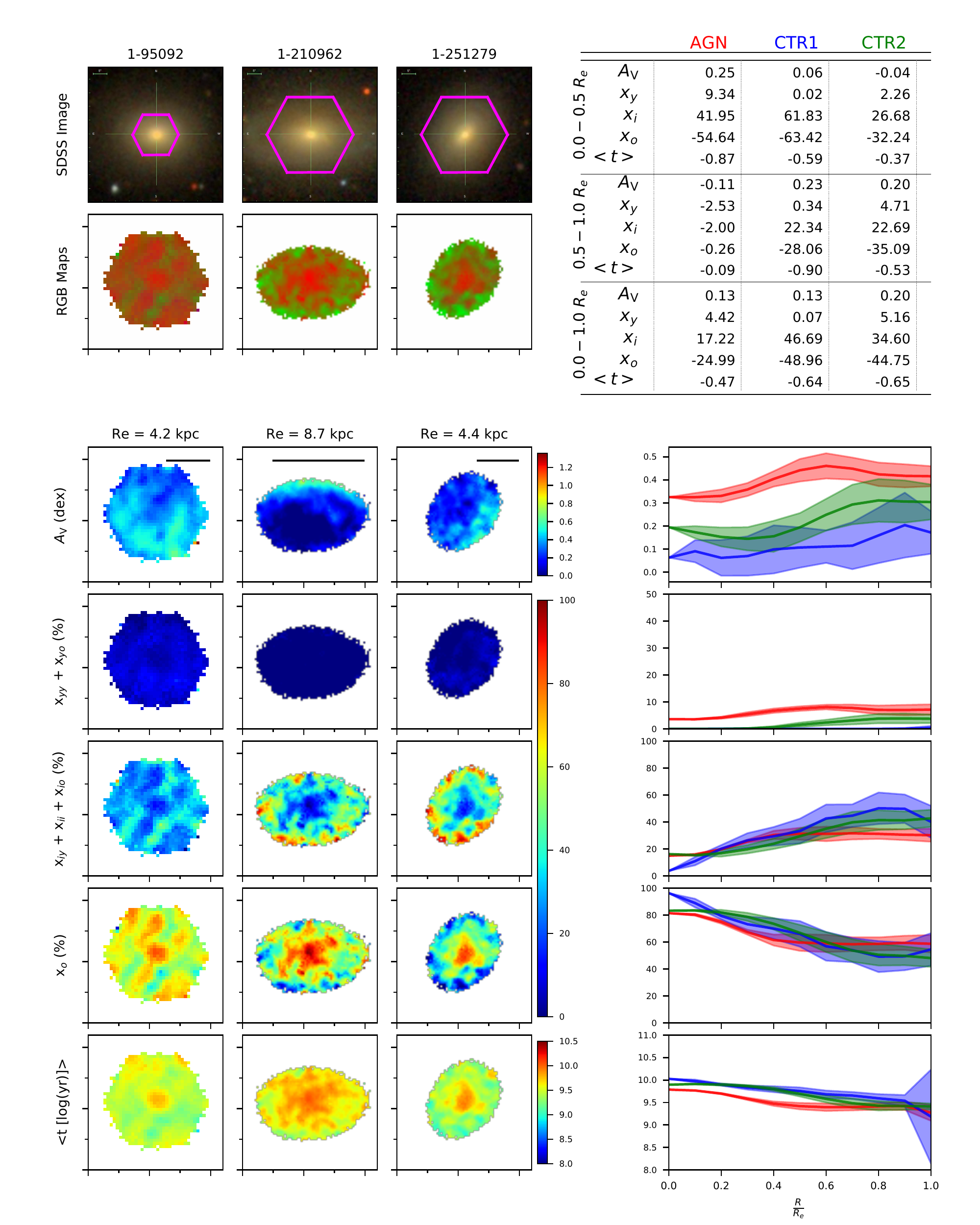}
    \caption{
    	Comparison between an early-type AGN and its control galaxies. See Fig. \ref{fig:late-comp} for description.
     }
    \label{fig:early-comp}
\end{figure*}

\begin{figure}
	\includegraphics[width=1.0\columnwidth]{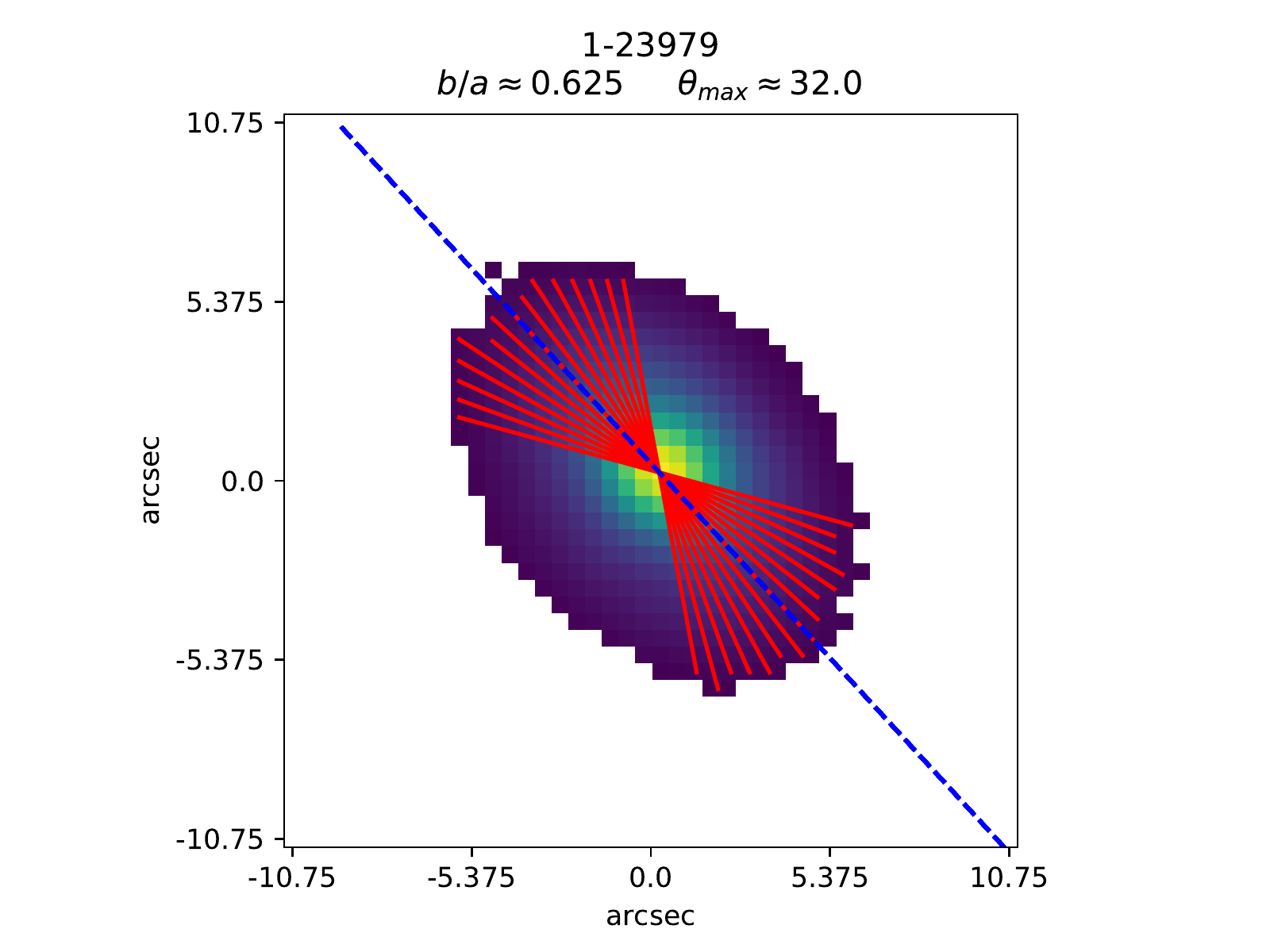}
    \caption{
    	Example of radial profile cuts on the normalization flux map of a galaxy (MaNGA ID: 1-23979) with $\approx$ 32 degrees of angle displacement from the major-axis (indicated with a blue dashed line). Red line segments are the radial profile cuts used to calculate the mean radial profile. Profiles closer to the minor axis were excluded to reduce any effects of projection (the more edge-on the galaxy is, the greater the projection distortion).
        }
    \label{fig:area_example}
\end{figure}

The base set used in the spectral synthesis is a reduced number of the SSPs calculated by \cite{Bruzual2003}. It comprises $N_\star$ = 46 elements (45 SSPs + 1 featureless continuum -- FC -- function of the form $F_\nu \propto \nu^{-1.5}$ to represent the AGN emission), spanning 15 ages (0.001, 0.003, 0.005, 0.010, 0.025, 0.040, 0.101, 0.286, 0.640, 0.905, 1.43, 2.50, 5.00, 11.00 and 13.00 Gyrs) and 3 metallicities (0.1, 1 and 2.5 $\rm Z_\odot$). The addition of a power law is necessary to account for the AGN continuum \citep{CidFernandes2004,Riffel2009}. For the foreground extinction, we used the \citet[CCM]{Cardelli1989} law, with $R_{\rm V} = 3.1$. The adopted normalization wavelength was $\lambda_0~=~5700$\,{\AA} and the synthesis was performed for the spectral range from $3800$\,{\AA} to $7000$\,{\AA}.


\subsection{Data Management}

In order to improve the management of the data which includes {\sc starlight} inputs and outputs, the compilation of the results and the analysis, we developed a software called {\sc megacube}. This software is designed to work with three main modules set up by a general configuration file. The modular approach was chosen with adaptability in mind, e.g., if a module was programmed to work with MaNGA datacubes' extraction, we could replace it with one that extracts another survey's datacubes. The modules used in this work are (in order of execution):

{\bf i) Data preparation: } This module is used to process and convert the MaNGA datacubes to a data format suitable for the chosen stellar population fitting code (\starlight). The main steps are as follows:

\begin{itemize}
\item Filtering of the spectra using a two-dimensional butterworth filter to remove spurious data (e.g., spiked values) and increase the signal to noise ratio, without combining adjacent spaxel, thus allowing to a better exploration of the spatial resolution. A better description of this technique  can be found in \cite{Riffel2016};
\item Galactic reddening correction of each spaxel using the Schlegel extinction maps \citep{Schlegel1998} and the CCM reddening law;
\item Redshift correction using the SDSS-III redshift provided in the {\it drpall} tables of the MaNGA database;
\item Estimation of the signal to noise ratio (SNR) in the wavelength range 5650-5750\,{\AA} for every spaxel;
\item Spaxels with SNR\,$<$\,$10$ were excluded when performing the fitting. This was done in order to have a good compromise between the spatial coverage and the reliability of the fitting results \citep[see][for details]{CidFernandes2004}.

\end{itemize}

{\bf ii) Spectral Fitting: } This module is used to invoke the fitting code and compile its results as described below:

\begin{itemize}
\item Setting up all the configuration files needed for the fits;
\item Fitting each individual spaxel with {\sc starlight};
\item Derivation of mean ages and metallicities, as well as star formation rates, from the starlight output;
\item Inclusion of both standard starlight output and derived parameters to the original datacubes as additional extensions.

\end{itemize}

{\bf iii) Analysis: } This module uses the fitting results to produce maps and radial plots (see Section \ref{res} for a better description):

\begin{itemize}
\item RGB maps: Qualitative representation of the spatially resolved stellar population age distribution for each galaxy, where the colors (red, green, and blue) represent three main age bins (see Sect. \ref{res});
\item Comparison figures: for each trio of galaxies (AGN and its two controls), panels showing the relevant properties (maps of stellar population properties derived from starlight and/or SDSS-III combined \emph{ugriz} images);
\item Radial profiles and gradients: for each galaxy and property, a mean profile (as well as its mean gradient value -- calculated as a function of R, dX/dR) is calculated to use for quantitative comparisons;
\item Gradients table: as result of the analysis, we have also generated a Table showing the gradients of the profiles for three different bins in terms of effective radius $R_e$;
\item {[\oiii]} $\lambda$5007 luminosity \loiii binned radial profiles: Radial profiles of the stellar population properties where the profiles for the AGN (and corresponding controls) are binned in groups according to the AGN luminosity. These profiles are shown also for AGN subsamples binned according to the host galaxy type: early and late-type.
\end{itemize}

It is worth mentioning that a similar organizer tool was developed by \citet{deAmorim2017} for the Calar Alto Legacy Integral Field Area (CALIFA) survey, which is a pioneer project of integral field spectroscopy legacy surveys.


\section{Results} \label{res}

The spectral synthesis gives as results, a number of output parameters, but we are mostly interested in $x_{j}$ -- the fractional contribution of each SSP to the total light at the normalization wavelength $\lambda_{0}$, that gives the SFH of each spaxel, and from which we also obtain the mean age for each spaxel, representing the age of the stellar populations as a single parameter. In addition, a valuable byproduct of the fitting is the amount of extinction in the line of sight, parameterized by the visual extinction $A_{\rm V}$. In Figs. \ref{fig:late-comp} and \ref{fig:early-comp} we illustrate the derived data products (maps, radial profiles and gradients) for two AGN and their respective control sample galaxies; the equivalent plots for the remaining of the sample is available in the Appendix.

In order to represent the galaxies' ages distribution with a single parameter at each spaxel, we calculated their light weighted mean age \citep{CidFernandes2005}, as follows:
 \begin{equation}
\langle log\,t_L\rangle \,= \frac{ \sum_{j=1}^{N_\star} x_j \, {\rm log}(t_j) }{ \sum_{j=1}^{N_\star} x_j },
\end{equation}
\noindent where t$_j$ is the age of the template j. The distributions of mean age are shown in the bottom row of the bottom left panels of Figs. \ref{fig:late-comp} and \ref{fig:early-comp}.

As stated by \citet{CidFernandes2005}, small differences in ages of individual SSPs are washed away in real data by noise effects. We therefore rebinned the population vectors in six stellar population components (SPCs): x$_{yy}$ ($1$\,Myr $\leq$ t $\leq 10$\,Myr), x$_{yo}$ ($10$\,Myr $<$ t $\leq$ $40$\,Myr), x$_{iy}$ ($40$\,Myr $<$ t $\leq$ $286$\,Myr), x$_{ii}$ ($286$\,Myr $<$ t $\leq$ $905$\,Myr), x$_{io}$ ($905$\,Myr $<$ t $\leq$ $2.5$\,Gyr), and x$_{o}$ ($2.5$\,Gyr $<$ t $\leq$ $13$\,Gyr).

We have also grouped the stellar population vector in three major age bins, described as follows:

\begin{itemize}
\item Young Age: $\rm X_Y = x_{yy} + x_{yo}$
\item Intermediate Age: $\rm X_I = x_{iy} + x_{ii} + x_{io}$
\item Old Age: $\rm X_O = x_{o}$
\end{itemize}

In order to visualize the spatial distribution of the populations' relative contributions in a qualitative way, RGB images of the galaxies were created by assigning the 3 colors (red, green, blue) to the binned population vectors: Red represents the old $\rm X_{O}$ ($2.5$\,Gyr $<$ t $\leq$ $13$\,Gyr), green the intermediate age $X_{I}$ ($40$\,Myr $ < t \leq 2.6$\,Gyr), and blue the young stellar populations $\rm X_{Y}$ ($1$\,Myr $\leq$ t $\leq 40$\,Myr).

In the bottom right of Figs. \ref{fig:late-comp} and \ref{fig:early-comp} we show mean radial profiles, up to $1.0\,R_e$, for each galaxy and property, derived using a nearest neighbor interpolation method to generate continuous values between pixel transitions and to remove abrupt changes due to spatially discrete maps. We opted for 30 equally spaced radial profiles in the galaxy plane limited to angular distances from the major axis of $\theta_{max} = {\rm tan}^{-1} (b/a)$ degrees, where $a$ and $b$ are, respectively, the semi-major and semi-minor axis of the SDSS galaxy image, obtained from the MaNGA's {\it drpall} table (calculated using S\'ersic profiles), as illustrated in Fig. \ref{fig:area_example}. The reason for this choice of maximum displacement from the major axis was the fact that profiles closer to the minor axis, when projected, resulted too noisy, possibly due to obscuration effects when the galaxies are too inclined relative to the line of sight. Mean profiles were then calculated for each property map by averaging all these profiles.

We have also calculated the mean gradients of each property (using the mean radial profiles) for 3 different regions: from $0.0$ to $0.5$ $R_e$, $0.5$ to $1.0$ $R_e$, and $0.0$ to $1.0$ $R_e$. Since we are comparing relatively low luminosity active galaxies with a matched control sample of non-active galaxies, the major differences should be detected in their nuclear regions. Thus, the division we used was decided based on qualitative observation of the average radial profiles, which revealed, in most galaxies, a trend of slope changes close to 0.5 $R_e$. Some values had to be extrapolated to a constant slope since the radial profiles could not reach 1.0 $R_e$ due to poor signal to noise ratio. These gradients for $A_{\rm V}$, $\rm X_{Y}$, $\rm X_{I}$ and $\rm X_{O}$, together with the gradients in mean age, are shown in a Table in the top right corner of Figs. \ref{fig:late-comp} and \ref{fig:early-comp}.

\subsection{AGN hosts {\it versus} control galaxies}

For each one of the AGN we show the radial variation of the derived properties compared to the two control galaxies in the Appendix. As an example, we have selected to show two typical sets of results in Figs. \ref{fig:late-comp} and \ref{fig:early-comp}, the results for one ``late-type" AGN (MaNGA ID 1-44379) and one ``early-type" AGN (MaNGA ID 1-95092) and their control galaxies.

In the case of the late-type AGN of Fig. \ref{fig:late-comp}, it shows that the RGB maps are dominated by the contribution of old populations in the centers for both the AGN hosts and the control galaxies. Just outward of the nucleus, younger populations dominate the AGN and the control galaxy CRT2 a bit further out, while the control galaxy CRT1 shows larger contribution of older age components also outside the nucleus. The extinction at the nucleus is stronger for the second control than the AGN while outwards they reach similar values that are higher than those of the first control. Regarding the contribution of the different stellar population age bins to the light at 5700\,\AA, there is no difference between the AGN and the controls for the youngest age bins, while for the intermediate age one, its contribution in the AGN is larger than in its control galaxies at all radii. There is also a difference for the old age bin $\rm X_{O}$, that is lower in the AGN than in the control galaxies everywhere inside $R_{e}$. These results also reflect in the mean age $<t>$: the mean age of the stellar population is lower everywhere in the galaxy for the AGN than for the control galaxies. The table listing the gradients also show differences between the AGN and control galaxies: for the inner radial bin ($0$-$0.5$ $R_{e}$) and for the full radial range ($0$-$1.0$ $R_{e}$), the AGN host galaxy shows steeper gradients than the controls for all properties.

The early-type AGN case of Fig. \ref{fig:early-comp} shows higher contribution of the old component in the central regions of the galaxies for the AGN and its controls. In this case, however, the AGN's RGB map shows a more homogeneous population throughout the galaxy, whilst the control galaxies' RGB maps show an older central region surrounded by a younger population outwards. The extinction for the AGN is larger compared with its control galaxies as can be seen in the $A_{\rm V}$ profiles. Although the $A_{\rm V}$ maps show higher values for the control galaxies, they are concentrated closer to the limits of $SNR \leq 10$, thus less reliable. The population profiles show a more constant distribution of ages for the AGN inside the $R \leq 1.0\,R_{e}$, specially between $0.5$ and $1.0\,R_{e}$. This behavior is reflected in the low gradient values of the $0.5-1.0\, R_{e}$ bin compared to the control galaxies. The mean age $<t>$ profile of the AGN shows a younger population over almost all of the $0-1.0\,R_{e}$ range, similar to the late-type case (Fig. \ref{fig:late-comp}. The gradients table shows that the outer region ($0.5-1.0\,R_e$) has a different behavior for the AGN compared to the control galaxies (which, in turn, behave similarly).

\begin{figure*}
	\includegraphics[width=2.\columnwidth]{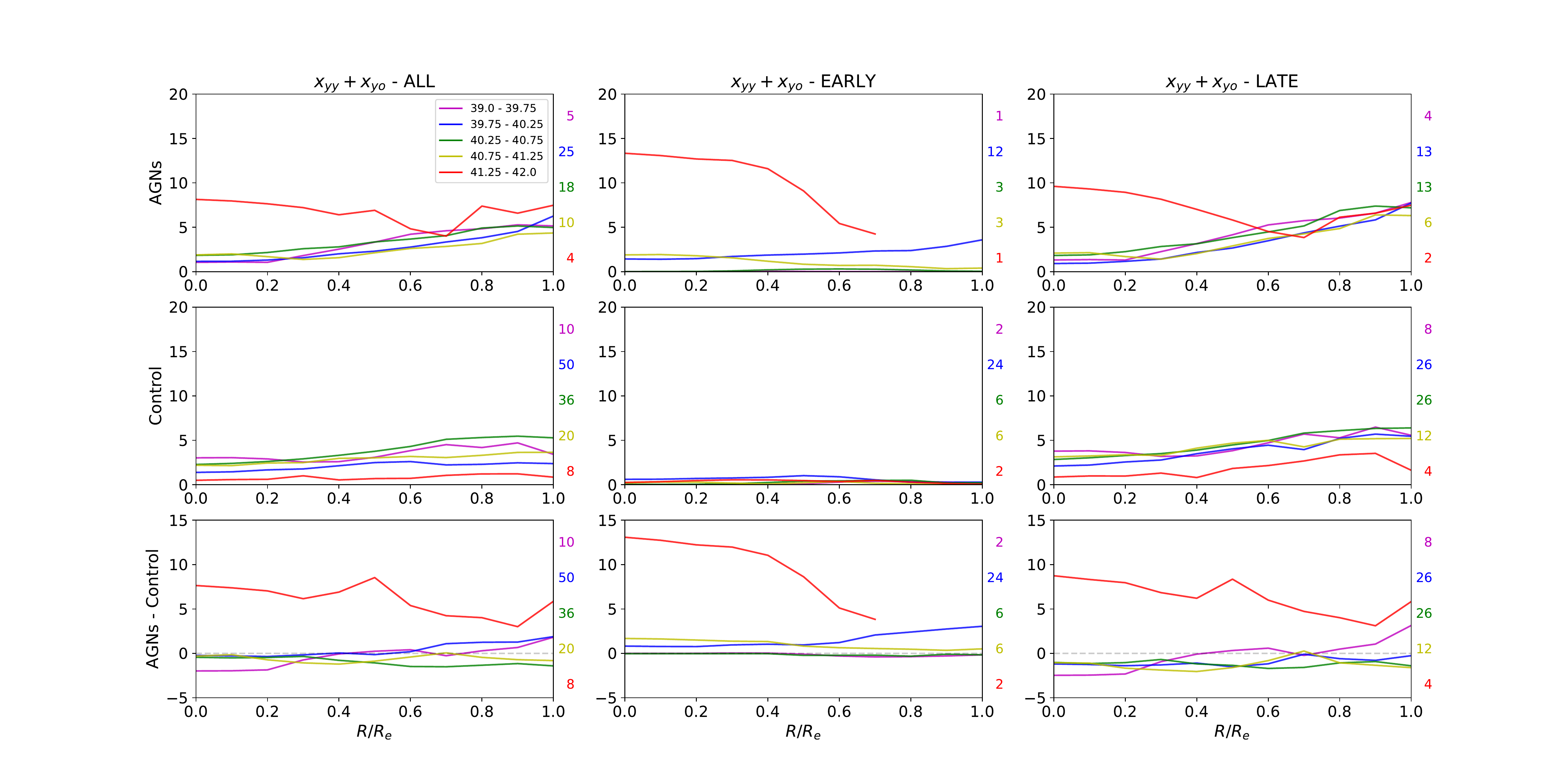}
    \caption{
        \loiii binned mean radial profiles for the {\bf x$_{yy}$} ($1$\,Myr $\leq$ t $\leq 10$\,Myr) and {\bf x$_{yo}$} ($10$\,Myr $<$ t $\leq$ $40$\,Myr) components combined. Each color pertains to the same \loiii range for every plot. The columns represent the groups of AGN used to calculate the average profiles for the AGN, its control galaxies, and their differences. The groups are, from left to right: all AGN, early-type AGN, late-type AGN. The rows, from top to bottom, show the average profiles for the AGN, the control galaxies (of the respective AGN group), and the differences. The colored numbers to the right of every plot are the quantity of galaxies used to calculate the mean profile of the same color.
        }
    \label{fig:global_xyy_xyo}
\end{figure*}

\begin{figure*}
	\includegraphics[width=2.\columnwidth]{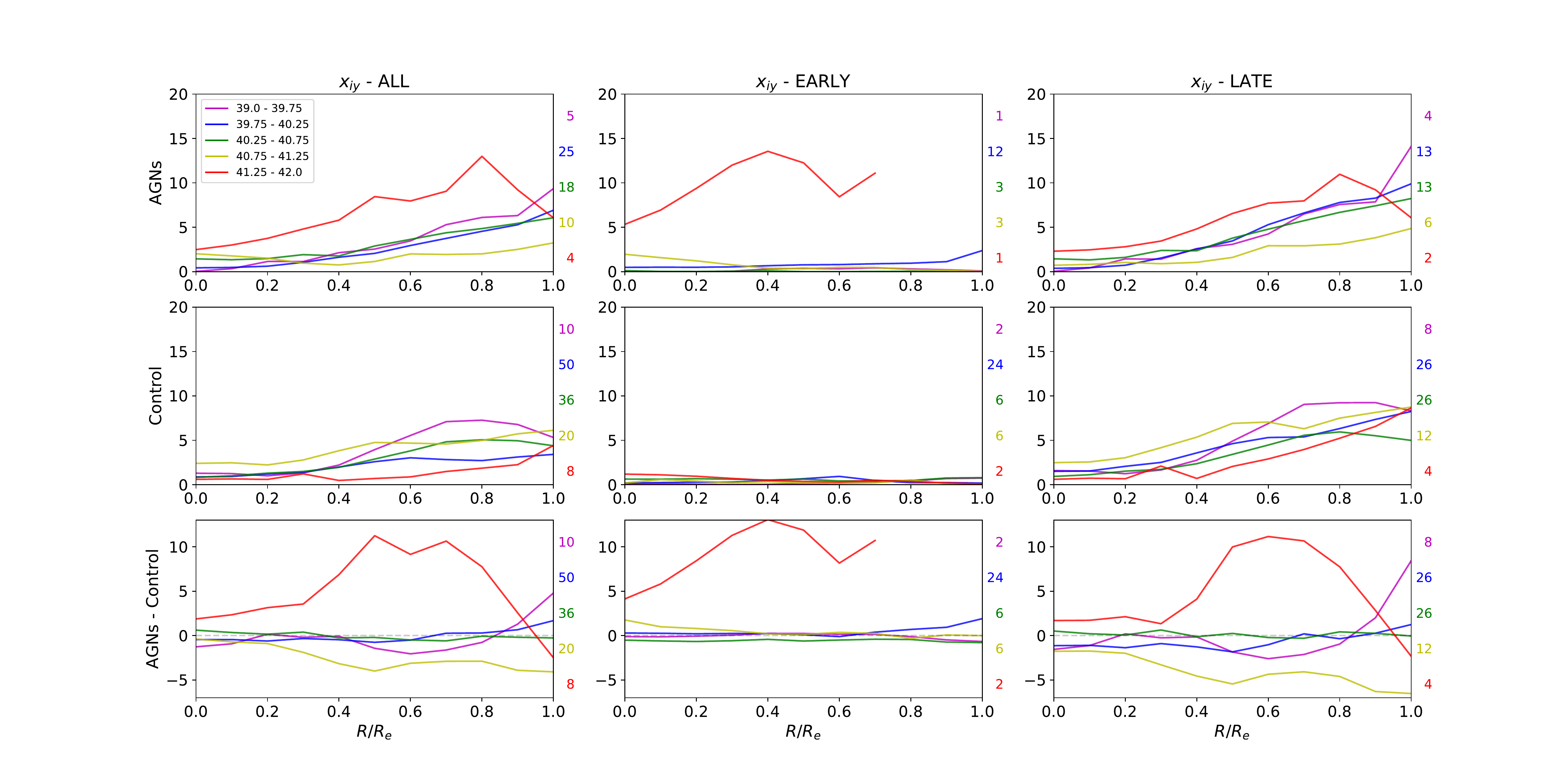}
    \caption{
        \loiii binned mean radial profiles for the {\bf x$_{iy}$} components ($40$\,Myr $<$ t $\leq$ $286$\,Myr). See figure \ref{fig:global_xyy_xyo} for the description.
        }.
    \label{fig:global_xiy}
\end{figure*}

\begin{figure*}
	\includegraphics[width=2.\columnwidth]{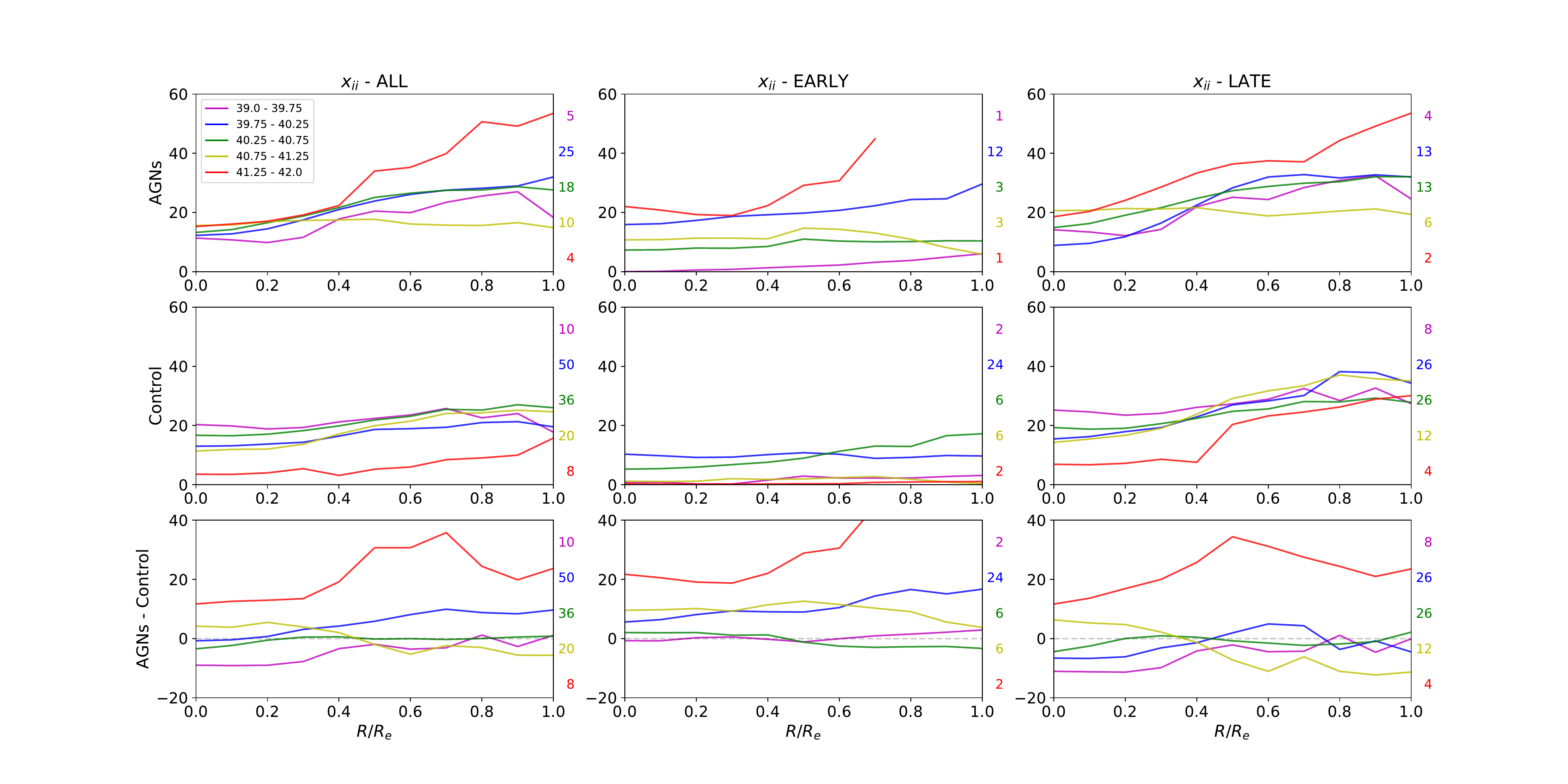}
    \caption{
        \loiii binned mean radial profiles for the {\bf x$_{ii}$} components ($286$\,Myr $<$ t $\leq$ $905$\,Myr). See figure \ref{fig:global_xyy_xyo} for the description.
        }
    \label{fig:global_xii}
\end{figure*}

\begin{figure*}
	\includegraphics[width=2.\columnwidth]{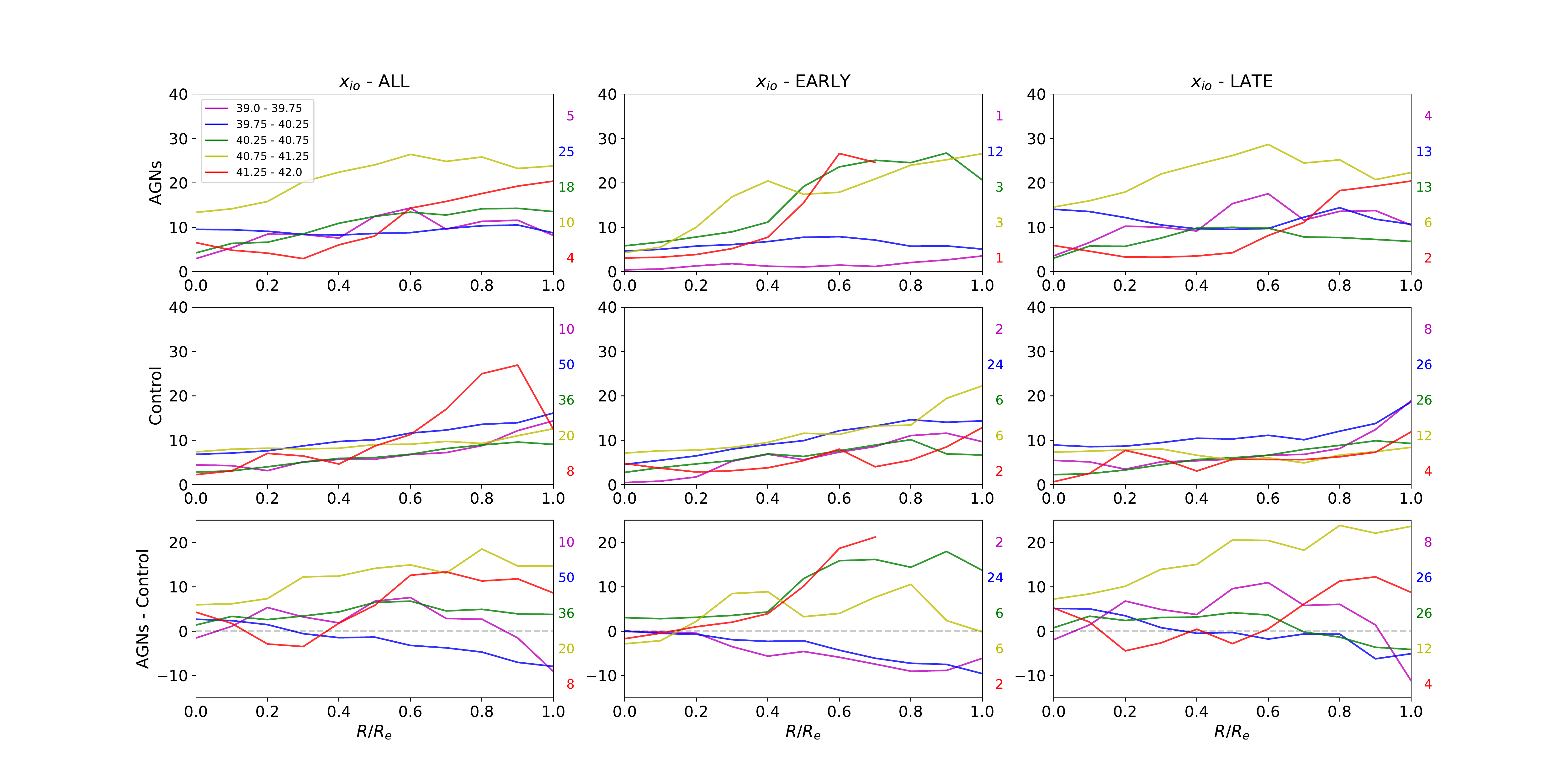}
    \caption{
        \loiii binned mean radial profiles for the {\bf x$_{io}$} components ($905$\,Myr $<$ t $\leq$ $2.5$\,Gyr). See figure \ref{fig:global_xyy_xyo} for the description.
        }
    \label{fig:global_xio}
\end{figure*}

\begin{figure*}
	\includegraphics[width=2.\columnwidth]{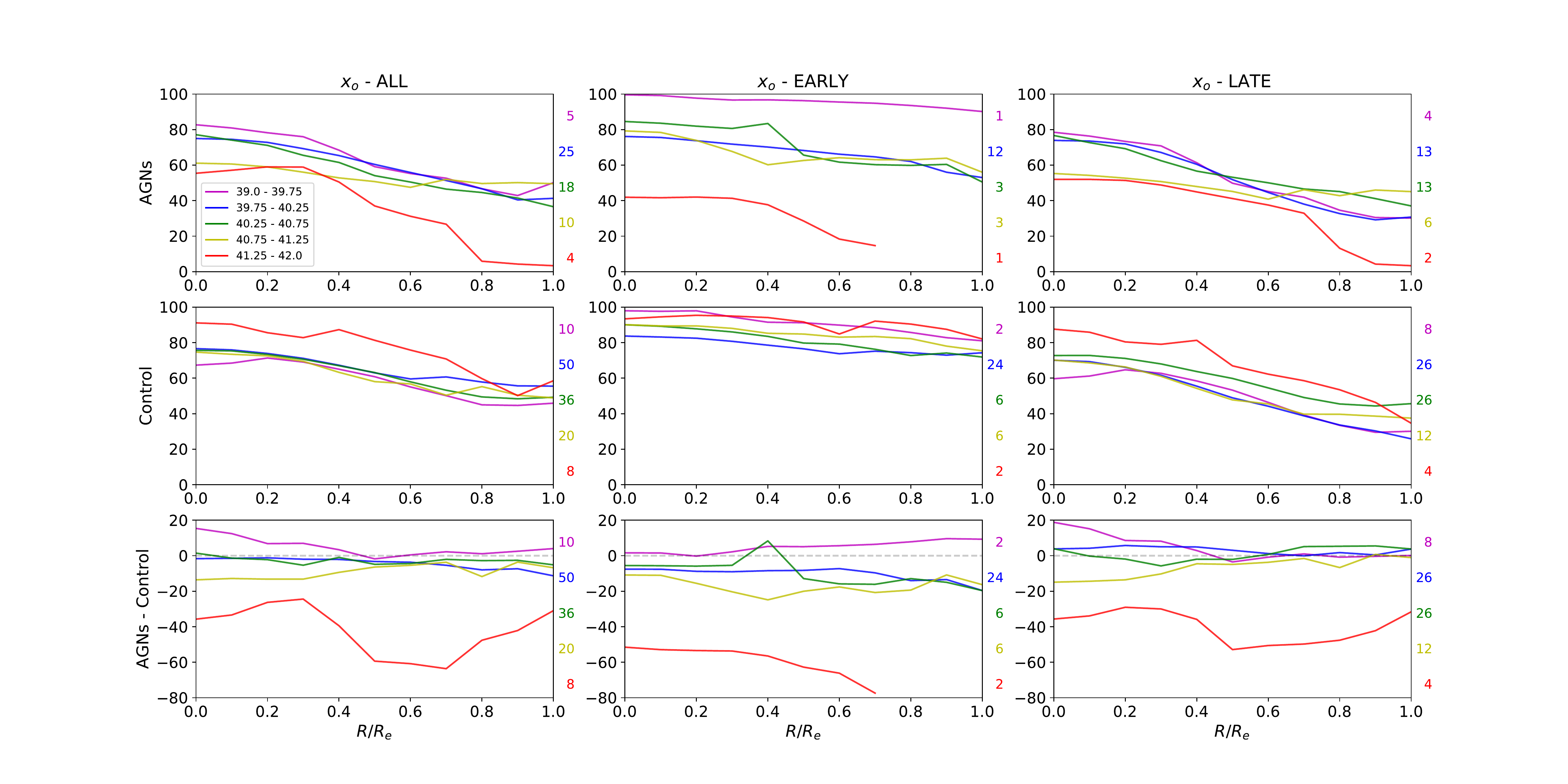}
    \caption{
        \loiii binned mean radial profiles for the {\bf x$_o$} components ($2.5$\,Gyr $<$ t $\leq$ $13$\,Gyr). See figure \ref{fig:global_xyy_xyo} for the description.
        }
    \label{fig:global_xo}
\end{figure*}

\begin{figure*}
	\includegraphics[width=2.\columnwidth]{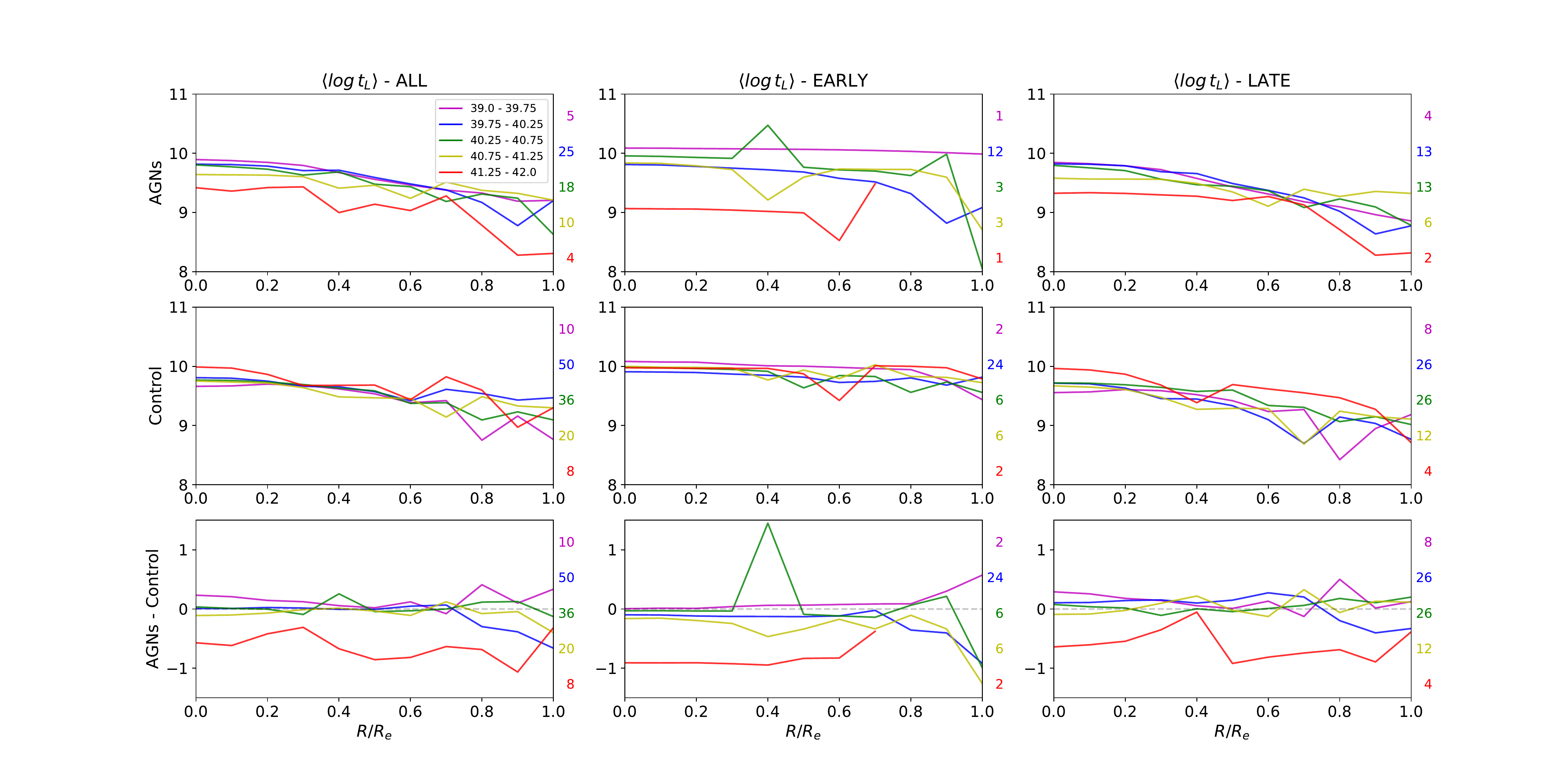}
    \caption{
        \loiii binned mean radial profiles for the {\bf mean age} ($\langle log\,t\rangle$). See figure \ref{fig:global_xyy_xyo} for the description.
        }
    \label{fig:global_m_ageL}
\end{figure*}


\begin{figure*}
	\includegraphics[width=2.\columnwidth]{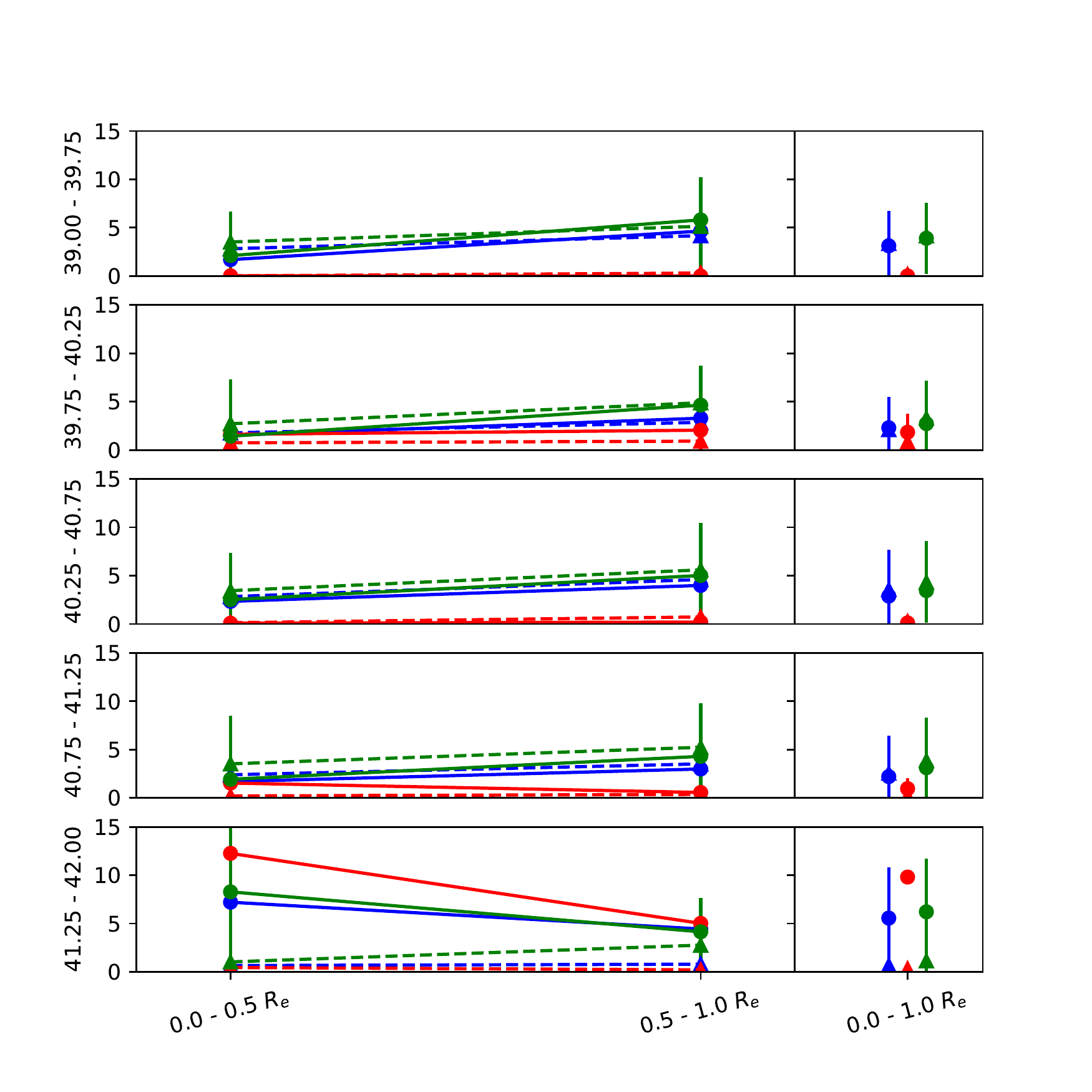}
    \caption{
        Young stellar population x$_y$ contribution for five different bins of luminosity ($39$-$39.75$, $39.75$-$40.25$, $40.25$-$40.75$, $40.75$-$41.25$, $41.25$-$42$), calculated for three different regions ($0.0$-$0.5\,R_e$, $0.5$-$1.0\,R_e$, $0.0$-$1.0\,R_e$). Each color represents a different AGN grouping: green for the late-type AGN, red for the early-type AGN, and blue for all the AGN sample. Solid lines correspond to the active galaxies and dashed lines to the control galaxies.
        }
    \label{fig:lum_young}
\end{figure*}

\begin{figure*}
	\includegraphics[width=2.\columnwidth]{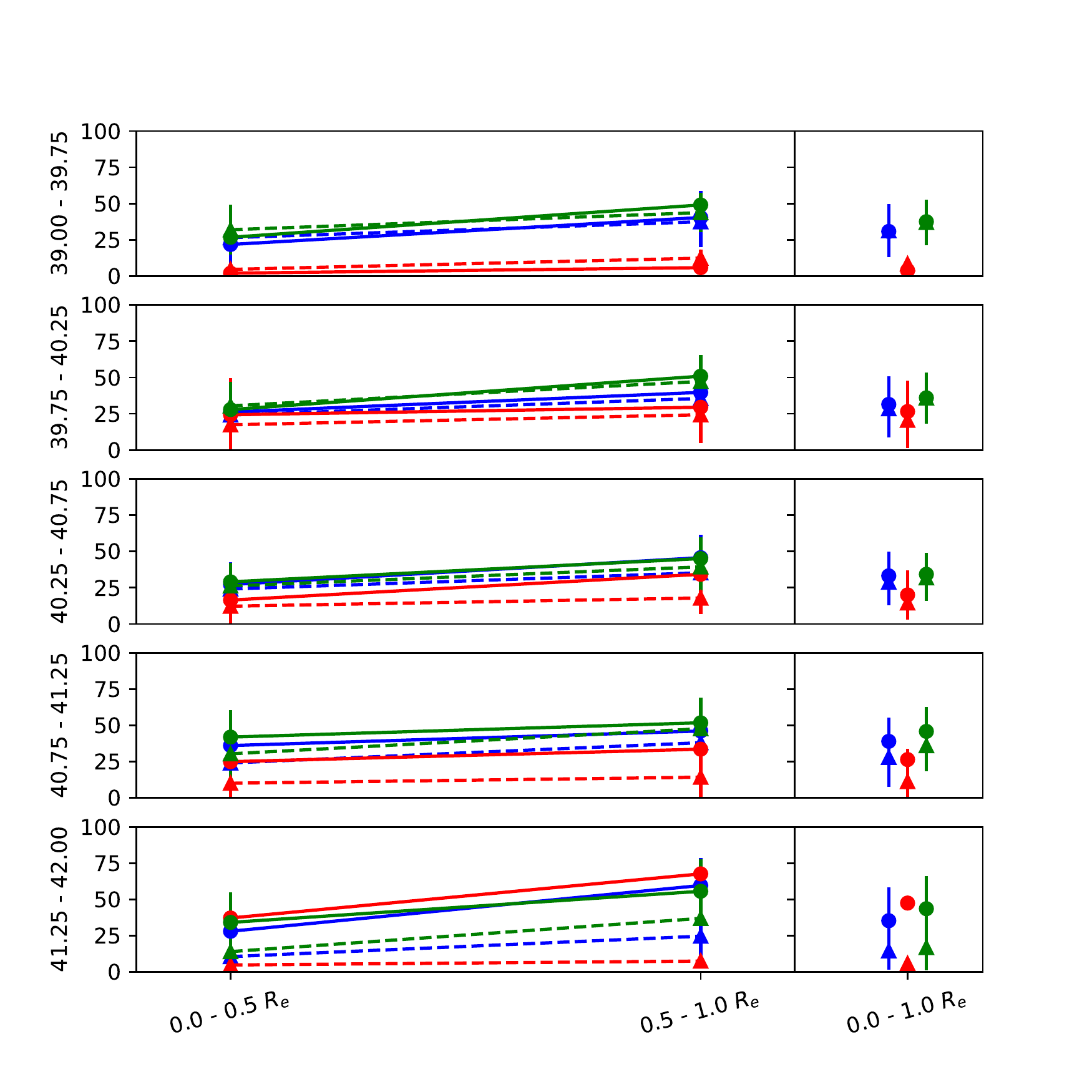}
    \caption{
        Intermediate age stellar population x$_i$ for different bins of luminosity, calculated for three different regions. See Fig. \ref{fig:lum_young}.
        }
    \label{fig:lum_interm}
\end{figure*}

\begin{figure*}
	\includegraphics[width=2.\columnwidth]{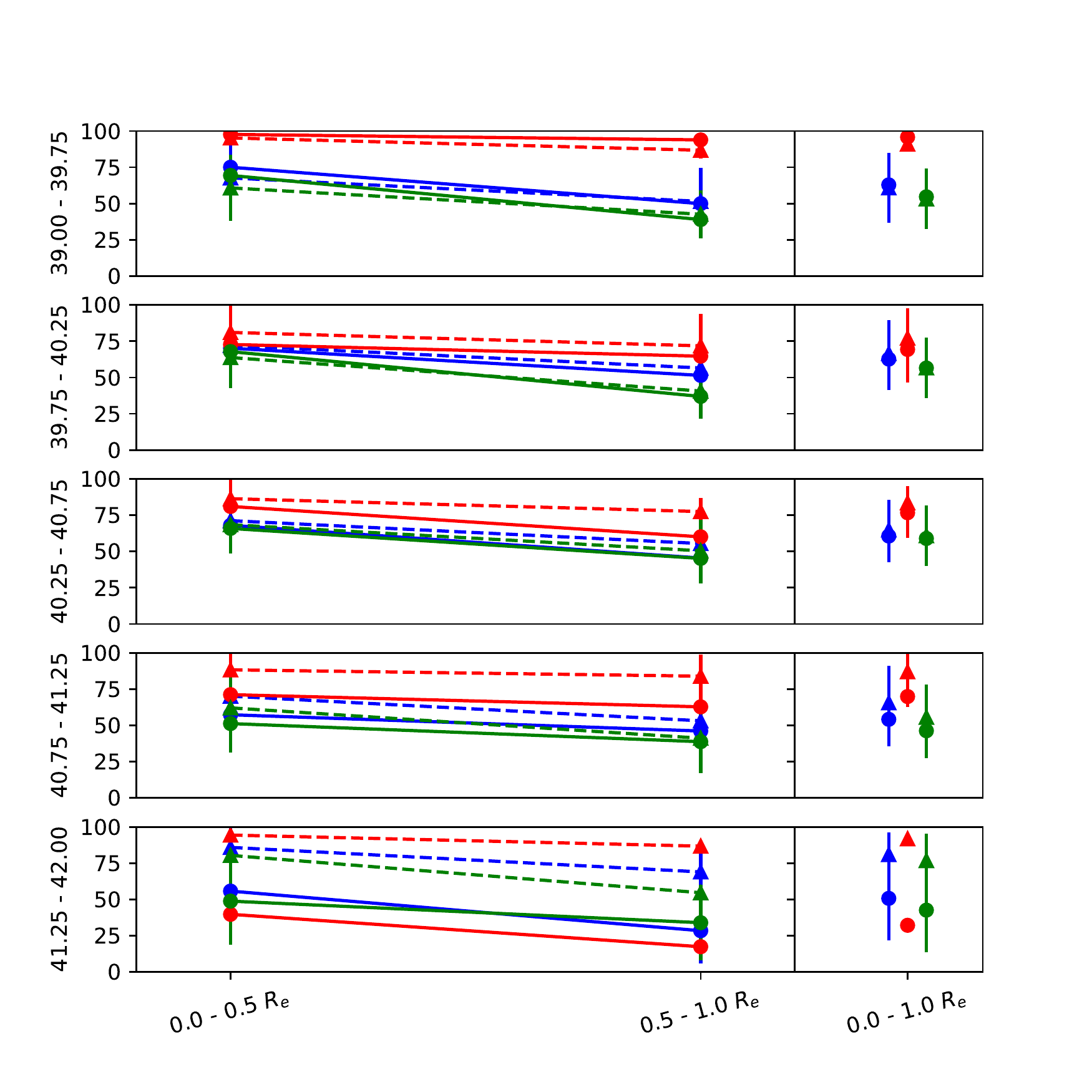}
    \caption{
        Old stellar population x$_o$ for different bins of luminosity, calculated for three different regions. See Fig. \ref{fig:lum_young}.
        }
    \label{fig:lum_old}
\end{figure*}

\begin{figure*}
	\includegraphics[width=1.8\columnwidth]{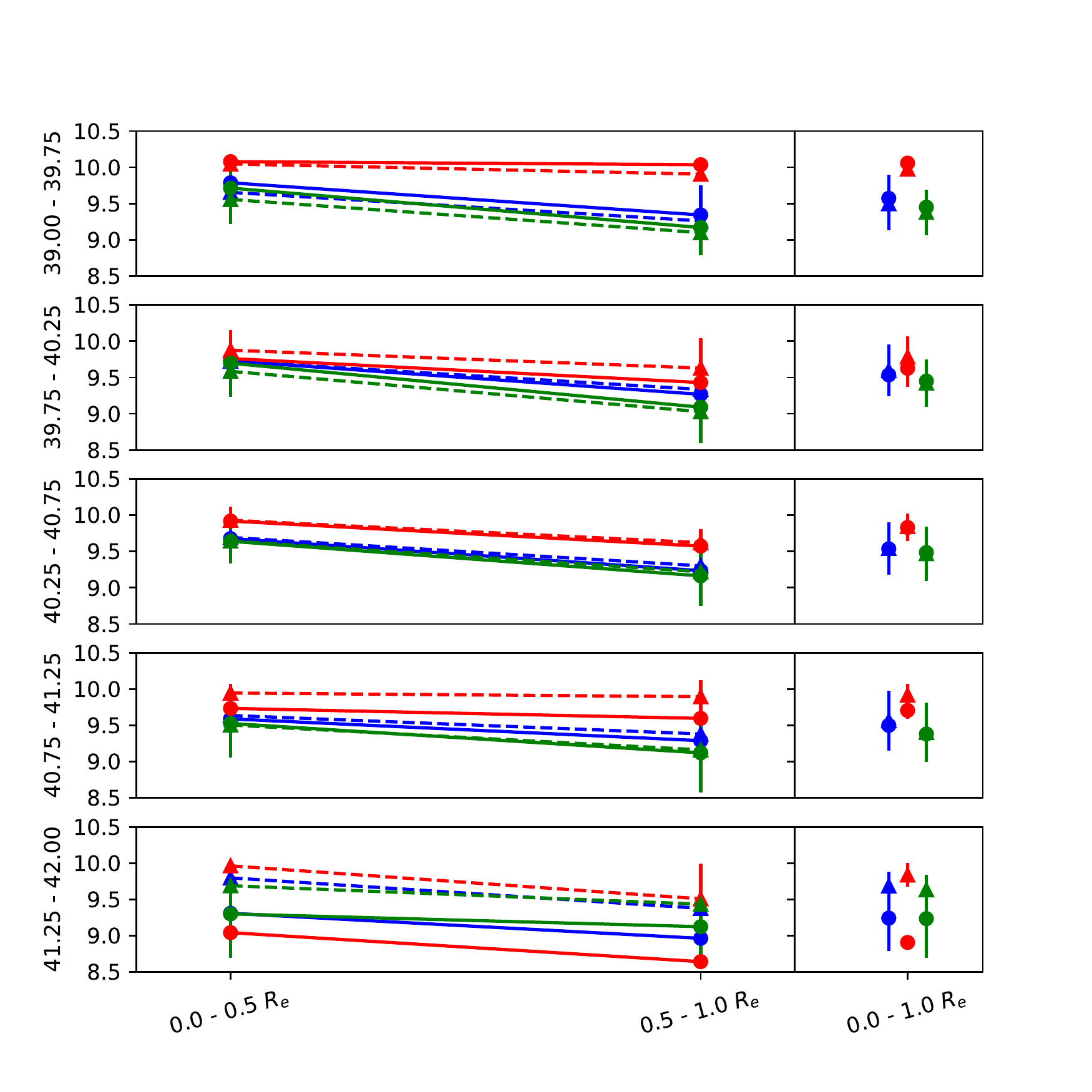}
    \caption{
        Mean age $\langle log\,t\rangle$ for different bins of luminosity, calculated for three different regions. See Fig. \ref{fig:lum_young}.
        }
    \label{fig:lum_m_ageL}
\end{figure*}

\begin{figure*}
	\includegraphics[width=1.8\columnwidth]{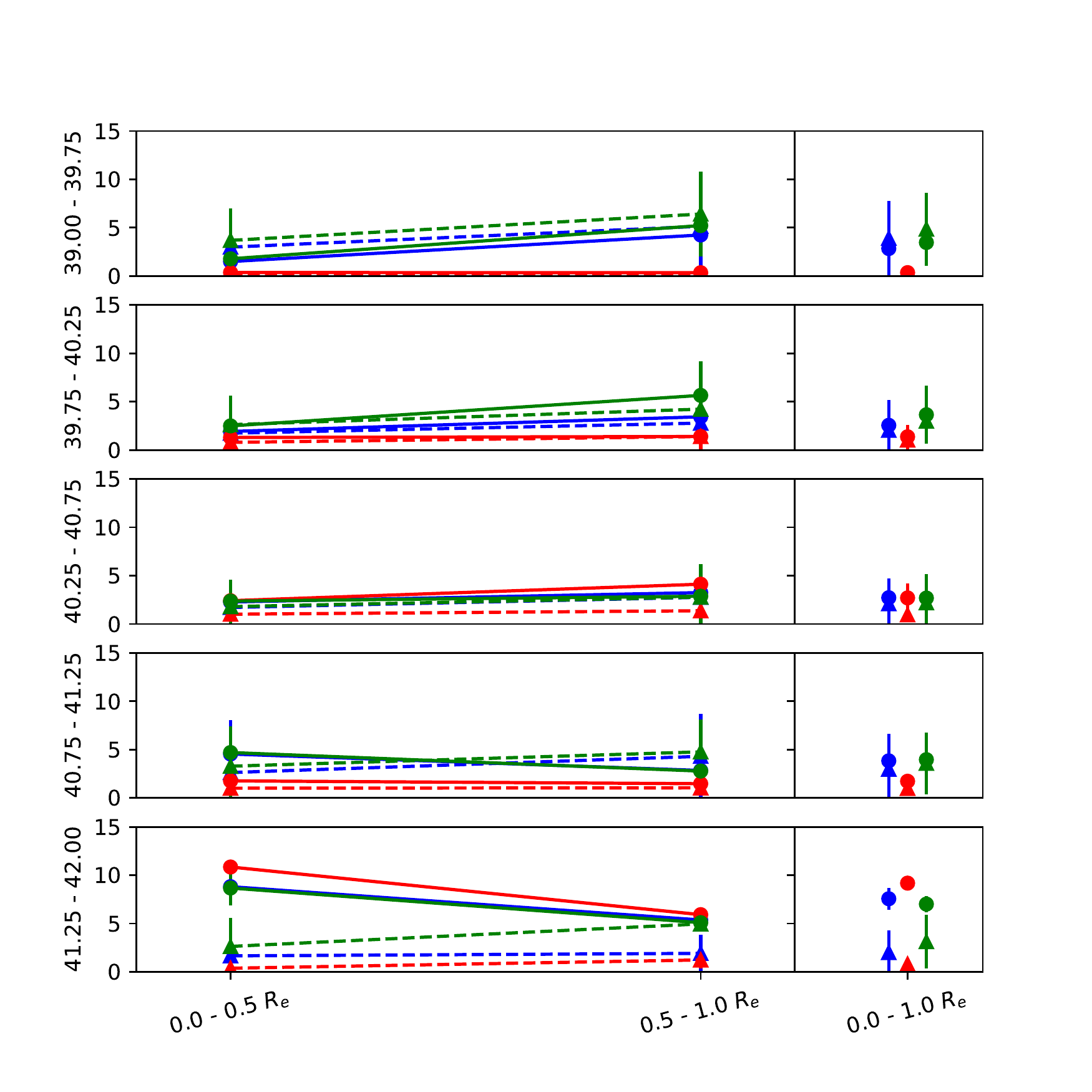}
    \caption{
        Featureless continuum for different bins of luminosity, calculated for three different regions. See Fig. \ref{fig:lum_young}.
        }
    \label{fig:lum_FC}
\end{figure*}


\section{Discussion} \label{disc}

In order to test for a possible relation between the AGN luminosity and the star formation history, we compare the stellar population profiles derived for the AGN and controls building average profiles for each of five L([\ion{O}{iii}]) bins. We grouped them in bins of $\rm log_{10}\,$ \loiii as follows: $39$ to $39.75$, $39.75$ to $40.25$, $40.25$ to $40.75$, $40.75$ to $41.25$, and $41.25$ to $42$, using the values for \loiii listed in Paper I.

In Figures \ref{fig:global_xyy_xyo} to \ref{fig:global_m_ageL} we show the mean radial profiles for the different age bins (see \S~\ref{res} as well as for the mean ages). In order to see if there are differences in these profiles for early and late type galaxies, we show the results both for all the galaxies grouped together and also separated in early and late-type hosts. In the top panels we show the results for the AGN hosts, in the middle panel for the corresponding controls and in the bottom panels the difference between AGN and controls. In each panel we show the five ranges of \loiii color coded and the respective number of objects (colored numbers) included in the average calculation. It is important to note that some of the AGN are not classified as early or late-type (because they could not be clearly classified, e.g. could be the result of mergers), so the sum of early and late-type galaxies may not represent the total number of objects (left column).

As can be seen in Fig. \ref{fig:global_xyy_xyo}, the clearest difference in the profiles occurs for the highest [\oiii] luminosity AGN, that show higher contribution of the young age component than the controls along the whole galaxy \footnote{It is worth mentioning that we have inspected carefully the fits for the highest luminosity AGN, where differences in the young population were detected, and no problems with the synthesis quality were found.}. In addition, the behavior seems not to depend on the type of host galaxy (early or late-type), at least up to $0.6 R_e$. We note that this bin has only 4 galaxies, but if we look at the profiles individually, we find the same behavior in each one of them. Comparing with the non-active galaxies, the corresponding profiles are similar to the other luminosity bins. For the late-type sources, there is a clear shift in the slope at $\sim 0.7 R_e$, for larger values of $\rm R_e$  it follows the other luminosity ranges. This, and the fact that the differences have a slightly negative slope in all morphology groups, suggests that the AGN may enhance the star formation process in the nuclear region ($\rm R_e \lesssim 0.6$).

A similar excess in the AGN as compared to controls is also observed in the intermediate age bins $x_{iy}$ and $x_{ii}$ shown in Figs. \ref{fig:global_xiy} and \ref{fig:global_xii}, although the excess seems to occur farther from the center. A comparison between Figs. \ref{fig:global_xyy_xyo}, \ref{fig:global_xiy} and \ref{fig:global_xii} suggests an age stratification with radius for the highest luminosity sources, showing a larger fraction of younger stars in the central region, with the intermediate age stars contributing more to the light in the outer regions. Regarding the old component, Fig. \ref{fig:global_xio} shows that, as expected, the high-luminosity AGN has less contribution of this component than the controls, with differences ranging from $\sim$20\% to 60\%. A small trend can be observed also in this figure, in the sense that a similar although smaller difference is also observed for the second luminosity bin, with the third and fourth bins showing almost no difference when compared with the controls, while the lowest luminosity bin showing the opposite: the controls having less contribution of the old component than the AGN.

In order to compare the galaxies using  a single parameter we plot the radial profiles of their mean ages (see \S\ref{res}) in Fig.~\ref{fig:global_m_ageL}. It is clear from this figure that the higher luminosity AGN hosts are younger than the control galaxies in the entire studied region (1$R_e$). We interpret these results as due to the fact that AGN hosts do prefer the ``outside-in" scenario for the recent star formation, while the galaxy's global stellar formation history (SFH) is better described by an inside out scenario.  This may reflect the idea that the active nuclei can drive the star formation process in the circumnuclear region.

Although the differences in the stellar population of the strongest AGN, when compared with the controls, were evident, a problem arises when trying to analyze other luminosity bins: the differences between the radial profiles are too noisy. In order to circumvent this problem, we binned the profiles into three regions ($0.0$-$0.5$ $R_e$, $0.5$-$1.0$ $R_e$, and $0.0$-$1.0$ $R_e$) and the stellar populations into $\rm X_{Y}$, $\rm X_{I}$, $\rm x_{O}$. We also plotted the contribution of the FC using these radial bins to inspect the light contribution of the AGN continuum. Figures \ref{fig:lum_young}-\ref{fig:lum_FC} show the results of this exercise\footnote{Some galaxies have limited $R_e$ coverage, meaning that the statistics gets weaker with radial distance. We addressed this problem by using radial intervals (0.5 $R_e$).}. What clearly emerges from this test is that the youngest SPs are concentrated in the inner $0.5\,R_e$ of the most luminous AGN hosts (Fig. \ref{fig:lum_young}) -- where the FC signature is the strongest (Fig. \ref{fig:lum_FC}) -- and intermediate age ones are located in regions with radius R $\gtrsim0.5\,R_e$ (see Fig.~\ref{fig:lum_interm}) for all AGN luminosities. In addition, the contribution of these stellar population components are much larger in the AGN hosts (circles) than that in the control galaxies (triangles) in the case of the highest luminosity AGN. Another result shown by these plots is that the youngest age contributions increase outwards for late-type galaxies (both controls and AGN -- except the most luminous AGN).

Figs. \ref{fig:lum_old} and \ref{fig:lum_m_ageL} show that the strongest (highest luminosity) AGN present, in general, younger SPs than their control objects. In addition these plots do allow us to better analyze the other luminosity bins. For log(L[\ion{O}{iii}] between 40.75 and 41.25 (the second most luminous bin) no significant differences can be seen for the younger populations, while the intermediate age contributions are higher for the AGN, being slightly more concentrated at the outer region (R $ \gtrsim 0.5\,R_e$). For the remaining luminosity bins a similar behavior is observed, however, when looking to the overall mean values (right side of Fig.~\ref{fig:lum_interm}) it is clear that the difference vanishes with luminosity decrease.  

In the case of the old stellar population bin, the contribution of this population in the AGN hosts is lower than that observed in the control objects, and a decrease is observed from the center outwards. A significant difference between active and non-active sources is seen for the two highest luminosity bins, specially when using the overall mean values (right side plot).  We also separate the objects according to their Hubble types (color coded) and no clear difference is observed between early and late-type galaxies when comparing active and non-active hosts.

The above results reinforce literature results \citep{Rembold2017,Kauffmann2003}, in the sense that when comparing low and high luminosity AGN, the contribution of old stellar populations decreases, while that of the younger stellar populations increases in the latter. However, our results do additionally show that this is specially enhanced in the circumnuclear regions  (R $\leq 0.5\,R_e$)  indicating that the the inflow of material feeding the AGN is partially being used to form stars. In addition, we suggest that these nuclear starbursts could at least be partially related to a positive AGN feedback, which may be inducing star formation in the host galaxy through enhancing the gas turbulence in the interstellar medium. Such a positive feedback is predicted by simulations \citep{Gaibler2012,Ishibashi2012,Wagner2012,Zubovas2013,Bieri2015,Zubovas2017} and was already detected in a few objects \citep{Cresci2015,Maiolino2017}.

As can be seen from Figs.\ref{fig:lum_young} to \ref{fig:lum_m_ageL} in general we observe that the fraction of young and intermediate age stellar populations increases with the radius, while in the case of the old population, it decreases. These results support the previous findings reported by \citet{Sanchez2013}, \citet{Ibarra-Medel2016} and \citet{Goddard2017} favoring an inside-out scenario for the formation of galaxies. However, when considering the most luminous AGN,  it no longer applies, and it seems that these AGN have been triggered by a recent supply of gas that has also triggered a recent star formation in their central regions. Our findings are opposite to the results of \citet{Goddard2017} in the case of early-type sources, we derive a slightly negative gradient while they derived a slightly positive one for this Hubble class\footnote{Note, however, that  we are studying the inner 1~R$_e $ (they used 1.5~R$_e$) and a smaller sample than that studied by these authors.}.  On the other hand, our findings seem to agree with those of \citet{Ibarra-Medel2016}, who showed that the radial stellar mass growth histories of early-type galaxies are on average nearly inside-out, though with a trend much less pronounced than that of the late-type galaxies.

\section{Conclusions}

We studied the stellar content of the first 62 AGN observed with SDSS-IV MaNGA and compared them with a matched sample of inactive galaxies presented in Paper~I. We constructed spatially resolved stellar population age maps, corresponding average radial profiles and gradients for these sources using the {\sc starlight} code, aimed at studying the effects of the AGN on the star formation history of the host galaxies. 

We found that the fraction of the young stellar population (t $\lesssim$ 40 Myr) is related with the AGN luminosity. For high-luminosity AGN (\loiii\ $\gtrsim 10^{41.25}$ ergs/s) it increases in the inner ($R \leq 0.5 R_e$) regions when compared with the objects in the control sample. In the case of the low-luminosity AGN, both AGN and control sample hosts, present very similar fractions of young stars. This result indicates that the inflow of material, besides feeding the nuclear engine, is being used to form new stars, thus rejuvenating the stellar content of the nuclear region of the AGN hosts. In addition, this very young starburst could also be enhanced by a positive AGN feedback produced by the high-luminosity AGN.

The fraction of the intermediate age, $X_{I}$ ($40$\,Myr $ < t \leq 2.6$\,Gyr), SP of the AGN hosts slightly increase outwards, with a clear enhancement over the entire galaxy when compared with the control sample. In addition, our results show that the inner region of the galaxies  are dominated by an old SP, whose fraction decreases outwards. These results support the previous findings of the CALIFA team \citep{Sanchez2013}, supporting an inside-out scenario for the galaxies' star formation history.

We also investigated for differences on the star formation histories between the different Hubble types. No significant differences were found between early and late-type hosts galaxies. 


From our results we suggest that an outside in scenario better describes the recent star formation in the AGN hosts, while an inside out scenario represents better the older generations of stars.

\section*{Acknowledgements}

We thank the anonymous referee for the useful comments and suggestions.
NDM thanks to CNPq for financial support. R.R. Thanks to FAPERGS and CNPq for financial support. 

Funding for the Sloan Digital Sky Survey IV has been provided by the Alfred P. Sloan Foundation, the U.S. Department of Energy Office of Science, and the Participating Institutions. SDSS acknowledges support and resources from the Center for High-Performance Computing at the University of Utah. The SDSS web site is www.sdss.org.

SDSS is managed by the Astrophysical Research Consortium for the Participating Institutions of the SDSS Collaboration including the Brazilian Participation Group, the Carnegie Institution for Science, Carnegie Mellon University, the Chilean Participation Group, the French Participation Group, Harvard-Smithsonian Center for Astrophysics, Instituto de Astrof\'isica de Canarias, The Johns Hopkins University, Kavli Institute for the Physics and Mathematics of the Universe (IPMU) / University of Tokyo, Lawrence Berkeley National Laboratory, Leibniz Institut f\"ur Astrophysik Potsdam (AIP), Max-Planck-Institut f\"ur Astronomie (MPIA Heidelberg), Max-Planck-Institut f\"ur Astrophysik (MPA Garching), Max-Planck-Institut f\"ur Extraterrestrische Physik (MPE), National Astronomical Observatories of China, New Mexico State University, New York University, University of Notre Dame, Observat\'orio Nacional / MCTI, The Ohio State University, Pennsylvania State University, Shanghai Astronomical Observatory, United Kingdom Participation Group, Universidad Nacional Aut\'onoma de M\'exico, University of Arizona, University of Colorado Boulder, University of Oxford, University of Portsmouth, University of Utah, University of Virginia, University of Washington, University of Wisconsin, Vanderbilt University, and Yale University.

\bibliographystyle{mnras}
\bibliography{references.bib}

\appendix

\section{AGN - Control Galaxies Comparison Images}

All the comparisons between AGN and its control galaxies will be available online.

\bsp	
\label{lastpage}
\end{document}